\documentclass[aps,twocolumn,showpacs,nofootinbib,footnote]{revtex4-1}

\usepackage{amsmath,amssymb,bm}
\usepackage[dvipdfmx]{graphicx}
\usepackage{hyperref}
\usepackage{yfonts}
\usepackage{color}
\newcommand{\beq}{\begin{eqnarray}}
\newcommand{\eeq}{\end{eqnarray}}

\newcommand{\U}{\text{U}}
\renewcommand\d{\partial}

\begin{document}

\title{Triangle anomalies and 
nonrelativistic Nambu-Goldstone modes of generalized global symmetries}

\author{Noriyuki Sogabe and Naoki Yamamoto}
\affiliation{Department of Physics, Keio University, Yokohama 223-8522, Japan}

\begin{abstract}
In massless QCD coupled to QED in an external magnetic field, a photon with the linear 
polarization in the direction of the external magnetic field mixes with the charge neutral 
pion through the triangle anomaly, leading to one gapless mode with the quadratic 
dispersion relation $\omega \sim k^2$ and one gapped mode. We show that this gapless 
mode can be interpreted as the so-called type-B Nambu-Goldstone (NG) mode 
associated with the spontaneous breaking of generalized global symmetries and that its 
presence is solely dictated by the anomalous commutator in the symmetry algebra.
We also argue a possible realization of such nonrelativistic NG modes in 3-dimensional 
Dirac semimetals.
\end{abstract}
\maketitle

\section{Introduction}
Spontaneous symmetry breaking (SSB) is one of the most important notions in modern 
physics that explains various physical phenomena from superfluidity to the origin of 
hadron masses. In general, when a continuous global symmetry is spontaneously broken, 
a gapless collective excitation called the Nambu-Goldstone (NG) mode emerges 
\cite{Nambu:1960tm, Nambu:1961tp, Goldstone:1961eq}. 
In a relativistic system with Lorentz invariance, the NG mode has the dispersion relation
$\omega = k$ and the number of NG modes, $N_{\rm NG}$, is equal to the number of 
broken symmetry generators, $N_{\rm BS}$ \cite{Goldstone:1962es}. This is not always the 
case in a system without Lorentz invariance, where the NG mode can have the dispersion 
$\omega \sim k^2$ and $N_{\rm NG}$ can be smaller than $N_{\rm BS}$ 
\cite{Nielsen:1975hm, Miransky:2001tw, Schafer:2001bq}. 

Whether an NG mode has the linear or quadratic dispersion relation is classified by the quantity 
$\rho_{ab} \equiv \langle [Q_a, Q_b] \rangle \neq 0$, with $Q_a$ broken symmetry generators 
and the expectation value taken in the vacuum: the NG mode characterized by 
nonvanishing $\rho_{ab}$ typically has the quadratic dispersion relation and is called type-B, 
while the rest typically has the linear dispersion relation and is called type-A. 
The generic counting rule for the numbers of the type-A and type-B NG modes, 
$N_{\rm A}$ and $N_{\rm B}$, is summarized as
\cite{Watanabe:2011ec, Watanabe:2012hr, Hidaka:2012ym}
\begin{align}
\label{NG}
N_{\rm NG} &= N_{\rm A} + N_{\rm B}, \\
\label{A}
N_{\rm A} &= N_{\rm BS} - 2N_{\rm B}, \\
\label{B}
N_{\rm B} &= \frac{1}{2}{\rm rank} \rho_{ab}\,.
\end{align}

So far, the notions of the SSB and NG modes have been mostly applied to the ordinary 
symmetries for pointlike objects. Yet, these notions can be generalized to symmetries 
for extended objects (such as strings and branes), called the generalized 
global symmetries or higher-form symmetries \cite{Gaiotto:2014kfa}.
In particular, the massless photon can be understood as a type-A NG mode associated 
with the SSB of a generalized global symmetry. 
More recently, it has been shown that photons with the quadratic dispersion relation 
(or nonrelativistic photons) appear in the axion electrodynamics with a spatially varying 
and periodic $\theta$ term and that it may be interpreted as a type-B NG mode \cite{Yamamoto:2015maz, Hidaka}. 
Such a situation is indeed expected to be realized in dense nuclear or quark matter 
in a magnetic field \cite{Yamamoto:2015maz, Brauner:2017mui} and a periodic array of 
topological and normal insulators \cite{Ozaki:2016vwu} (see also Ref.~\cite{Qiu:2016hzd}).

In this paper, we show that the emergence of the type-B NG mode of generalized global 
symmetries is more generic than previously thought and it appears in a much more simple 
setup: strongly interacting massless Dirac fermions coupled to a dynamical U(1) gauge field 
in a background magnetic field. One example is QCD coupled to QED in the external 
magnetic field. In particular, translational symmetry breaking by the spatially varying $\theta$ 
term in the previous examples \cite{Yamamoto:2015maz, Brauner:2017mui, Ozaki:2016vwu}, 
which was conjectured in Ref.~\cite{Brauner:2017mui} to be the only way to realize the 
nonrelativistic photon, is not necessary in this case.

Our main purpose of this paper is to show further that the presence of this type-B NG mode is 
solely dictated by the anomalous commutator related to the triangle anomaly in the symmetry 
algebra. In this sense, this phenomenon is universal depending on the SSB of generalized global 
symmetries, SSB of chiral symmetry, and the presence of the triangle anomaly. As we shall 
also discuss, this type-B NG mode should be experimentally testable, e.g., in Dirac semimetals 
in 3-dimensional solids.

\section{Example}
\subsection{Setup}
As an example, we consider massless two-flavor QCD coupled to QED in an external 
homogeneous magnetic field ${\bm B}_{\rm ex}$. A similar setup with finite quark mass was 
previously studied in Ref.~\cite{Brauner:2017uiu}.
The presence of the external magnetic field explicitly breaks chiral symmetry down to 
$\U(1)^{\tau_3}_{\rm V} \times \U(1)^{\tau_3}_{\rm A}$, which is the invariance under the transformation 
$q \rightarrow {\rm e}^{{\rm i}\alpha_{\rm V}\tau_3} {\rm e}^{{\rm i}\alpha_{\rm A}\tau_3 \gamma_5}q$, 
with $\tau^3$ being one of the SU(2) generators.
This partial chiral symmetry is spontaneously broken to $\U(1)^{\tau_3}_{\rm V}$ in the vacuum.
The low-energy degrees of freedom below the energy scale of charged pions are the charge 
neutral pion $\pi^0$ and massless photons, which we shall focus on below.

At the leading order in derivatives, the effective Lagrangian for $\pi^0$ and electromagnetic field 
$F^{\mu \nu}$ is
\beq
\label{L}
{\cal L} = \frac{1}{2}\d_{\mu} \pi^0 \d^{\mu} \pi^0 -\frac{1}{4} F_{\mu \nu} F^{\mu \nu}
- \frac{C}{4} \frac{\pi^0}{f_{\pi}} F_{\mu \nu} \tilde F^{\mu \nu},
\eeq
where the last term is the Wess-Zumino-Witten (WZW) term \cite{Wess:1971yu, Witten:1983tw}
that accounts for the triangle anomaly in this low-energy effective theory with $C=1/(4\pi^2)$, 
$f_{\pi}$ the pion decay constant, and 
$\tilde F^{\mu \nu} = \frac{1}{2}\epsilon^{\mu \nu \alpha \beta}F_{\alpha \beta}$.
(The factor 1/4 in front of the WZW is just for later convenience.) 
We assume the local charge neutrality in the vacuum. 

Let us consider the fluctuations of $E^i = -F^{0i}$, $B^i = -\tilde F^{0i}$, and $\pi^0$ 
around the QCD vacuum in the background magnetic field. We set 
${\bm B} = {\bm B}_{\rm ex} + \delta {\bm B}$, ${\bm E} = \delta{\bm E}$, and $\pi^0 = \delta \pi^0$
and we focus on the first order in fluctuations.
The equations of motion for $\pi^0$ and electromagnetic fields are 
\begin{gather}
\label{EOM_pi}
(\d_t^2 - {\bm \nabla}^2) \delta \pi^0 = \frac{C}{f_{\pi}} {\bm B}_{\rm ex} \cdot \delta {\bm E}\,, \\
\label{Gauss}
{\bm \nabla} \cdot \delta {\bm E} = -\frac{C}{f_{\pi}} {\bm \nabla} \delta \pi^0 \cdot {\bm B}_{\rm ex}\,,
\\
\label{Ampere}
-\d_t \delta {\bm E} + {\bm \nabla} \times \delta {\bm B} = \frac{C}{f_{\pi}} \d_t \delta \pi^0 {\bm B}_{\rm ex}\,,
\end{gather}
respectively. 
The right-hand sides of Eqs.~(\ref{Gauss}) and (\ref{Ampere}) are the anomalous charge 
and current induced by the pion fluctuation. In particular, the latter is the counterpart of 
the chiral magnetic effect (CME) in the hadronic phase \cite{Yamamoto:2015maz} 
(see also Ref.~\cite{Fukushima:2012fg}). 
Indeed, if one identifies $\d_t \delta \pi^0/f_{\pi}$ with the ``chirality imbalance" of chiral 
fermions defined by $\mu_{\rm R} - \mu_{\rm L} \equiv 2\mu_5$, it exactly takes the form 
of the CME given, e.g., in Ref.~\cite{Fukushima:2008xe}. 
This is similar to the fact that $\d_t \theta$ plays the role of the chirality imbalance in 
the axion electrodynamics \cite{Kharzeev:2009fn}. Note that the anomalous Hall effect
of the form $(C/f_{\pi}) {\bm \nabla} \delta \pi^0 \times \delta{\bm E}$ \cite{Yamamoto:2015maz} 
is higher order in fluctuations and is ignored in the right-hand side of Eq.~(\ref{Ampere}).

Taking the time derivative of Eq.~(\ref{Ampere}), and then using the Faraday's law 
$\d_t {\bm B} = - {\bm \nabla} \times {\bm E}$ and Eq.~(\ref{Gauss}), we obtain
\beq
\label{EOM_E}
(\d_t^2 - {\bm \nabla}^2) \delta{\bm E} =\frac{C}{f_{\pi}} 
\left[{\bm \nabla} ({\bm \nabla} \delta \pi^0 \cdot {\bm B}_{\rm ex})
-\d^2_t \delta \pi^0 {\bm B}_{\rm ex} \right]\,.
\eeq
It is clear from Eqs.~(\ref{EOM_pi}) and (\ref{EOM_E}) that, when $\delta {\bm E}$ and 
${\bm \nabla} \delta \pi^0$ are perpendicular to ${\bm B}_{\rm ex}$, 
the photon does not receive any anomalous correction expressed by the right-hand sides, 
and the dispersion relation is just given by the usual one $\omega = k$. On the other hand, 
when $\delta {\bm E} \cdot {\bm B}_{\rm ex} \neq 0$ and/or ${\bm \nabla} \delta \pi^0 \cdot {\bm B}_{\rm ex} \neq 0$, 
the behavior of the photon is qualitatively modified by the anomalous effects.

\subsection{Dispersion relations}
Without loss of generality, we take the direction of the external magnetic field in the 
$z$ direction and set ${\bm B}_{\rm ex} = B_{\rm ex} \hat {\bm z}$. 
As the simplest case, we assume that $\pi^0$ and photon are propagating in the 
$x$ direction; we also assume that the linear polarization $\delta {\bm E}$ is along 
the $z$ direction, $\delta {\bm E} = \delta E_z \hat {\bm z}$.
(From the argument above, the other linear polarization in the $y$ direction does not 
receive any correction.) The extension to more generic directions is straightforward.
We look for the solutions of the form 
$\delta \pi^0 \propto {\rm e}^{-{\rm i}\omega t + {\rm i}k x}$ and
$\delta E_z \propto {\rm e}^{-{\rm i}\omega t + {\rm i}k x}$.
From Eqs.~(\ref{EOM_pi}) and (\ref{EOM_E}), the equations of motion for 
$\delta \pi^0$ and $\delta E_z$ in momentum space can be summarized in the matrix 
equation $D_{i j} a^j =0$, where
\beq
D \equiv \!
\left(
\begin{tabular}{cc}
$\omega^2 - k^2$ & $\alpha$  \\
$\alpha \omega^2$ & $\omega^2 - k^2$  \\
\end{tabular}
\right)\,\!,
\quad \! \!
{\bm a} \equiv \! \left(
\begin{tabular}{c}
$\delta \pi^0$ \\
$\delta E_z$ 
\end{tabular}
\right)\,\!.
\eeq
Here we defined
\beq
\alpha \equiv \frac{C B_{\rm ex}}{f_{\pi}}\,.
\eeq

We can derive the particle spectrum and dispersion relations by the diagonalization of $D$.
As a result, we obtain one gapless mode with the quadratic dispersion relation and one
gapped mode,
\beq
\label{dispersion}
\omega = \frac{k^2}{\alpha} + O(k^4)\,, \qquad \quad \omega = \alpha + O(k^2)\,,
\eeq
respectively. 

Without the external magnetic field, we originally have three gapless modes---photon with 
two polarizations and $\pi^0$. This result shows that the mixing between the photon with 
one of the polarizations and $\pi^0$ due to the WZW term leads to one gapless mode 
with the quadratic dispersion relation (while the photon with the other polarization remains 
unchanged and has the linear dispersion relation). 
This phenomenon looks seemingly similar to the emergence of the type B NG modes, 
whose previous examples include magnons in ferromagnets and NG modes in relativistic 
Bose-Einstein condensation \cite{Miransky:2001tw, Schafer:2001bq}.

\subsection{Temporal gauge}
It is easy to show that the nonrelativistic gapless mode in Eq.~(\ref{dispersion}) can also be 
understood as a type-B NG mode in a particular choice of gauge---the temporal gauge 
$A_t =0$. In this gauge, the WZW term in Eq.~(\ref{L}) to the second order in fluctuations 
reduces to
\beq
\label{mix}
{\cal L}_{\rm mix} = \alpha \pi^0 \d_t A_z.
\eeq
At sufficiently low energy, the kinetic terms for $\pi^0$ and $F^{\mu \nu}$ are negligible 
compared with this mixing term. In this regime, we can see from Eq.~(\ref{mix}) that 
$A_z$ and $\pi^0$ are canonically conjugate degrees of freedom, and the corresponding 
NG mode must be just one, but not two, and hence, it is the type-B mode. 

Note here that, without the background magnetic field, the leading term in the WZW term
is third order in fluctuations of ${\bm E}$, ${\bm B}$ and $\pi^0$, and then $A_z$ and 
$\pi^0$ would not be canonically conjugate. This may simply be understood from the 
fact that a type-B NG mode could not appear in the absence of the explicit breaking 
of Lorentz invariance when the background magnetic field is turned off.

Note also that the structure of Eq.~(\ref{mix}) is similar to the Lagrangians considered 
in Refs.~\cite{Miransky:2001tw, Schafer:2001bq, Nambu:2004yia}. 
However, the nonrelativistic gapless mode here is distinct from the previous examples of 
type-B NG modes in that this is an NG mode not associated with a usual symmetry for 
pointlike objects, but with a generalized global symmetry for extended objects.

\section{Symmetry algebra}
\subsection{Photon as a type-A NG mode of 1-form symmetries}
Before considering the relation of the nonrelativistic gapless mode in Eq.~(\ref{dispersion}) 
to the SSB of a 1-form symmetry, we first recall, following Ref.~\cite{Gaiotto:2014kfa}, that 
the usual photon with the dispersion relation $\omega = k$ can be interpreted as a type-A 
NG mode associated with the SSB of 1-form symmetries.

In the vacuum without $\pi^0$ and ${\bm B}_{\rm ex}$, we have 
\begin{align}
\label{cons_E}
\d_{\mu}F^{\mu \nu} &= 0, \\
\label{cons_B}
\d_{\mu} \tilde F^{\mu \nu} &= 0,
\end{align}
where Eq.~(\ref{cons_E}) follows from the equation of motion in the vacuum and 
Eq.~(\ref{cons_B}) follows from the Bianchi identity. These two equations can be regarded 
as the conservation laws for the 2-form currents, $j_{\rm E}^{\mu \nu} \equiv F^{\mu \nu}$ and 
$j_{\rm M}^{\mu \nu} \equiv \tilde F^{\mu \nu}$, which physically stand for the conservation 
laws of the electric and magnetic fluxes through a codimension-2 surface. 
They are the consequences of the 1-form symmetries, 
$A_{\mu} \rightarrow A_{\mu} + \lambda_{\mu}$ (electric symmetry) 
and $\tilde A_{\mu} \rightarrow \tilde A_{\mu} + \tilde \lambda_{\mu}$ (magnetic symmetry) 
in Maxwell theory without matter, where $\tilde F_{\mu \nu} = \d_{\mu} \tilde A_{\nu} - \d_{\nu} \tilde A_{\mu}$,
and $\lambda$ and $\tilde \lambda$ are flat connections.

The operators charged under the 1-form electric and magnetic symmetries are the 
Wilson loop and 't Hooft loop,
\beq
W(C) \equiv \exp \left({\rm i}\int_C A\right), \quad
T(C) \equiv \exp \left({\rm i}\int_{C} \tilde A\right),
\eeq
respectively, where $C$ is a 1-dimensional closed loop. One can easily show that, 
in the vacuum (Coulomb phase), $\langle W(C) \rangle \neq 0$ and $\langle T(C) \rangle \neq 0$ 
under proper normalizations \cite{Gaiotto:2014kfa}, and both the electric and magnetic symmetries 
are spontaneously broken. We then have
\begin{align}
\label{EW}
\int_{\Sigma} \langle [j_{\rm E}^{0i}({\bm x}), W(C)] \rangle \neq 0, \quad
\int_{\Sigma} \langle [j_{\rm M}^{0i}({\bm x}), T(C)] \rangle \neq 0, 
\end{align}
where the integral is taken over the codimension-2 surface $\Sigma$ normal to the $i$ direction. 
Equation (\ref{EW}) suggests that the total number of broken symmetry generators is six.
However, not all of them are independent because of the commutation relation,
\beq
\label{EB}
[E^i({\bm x}), B^j(\bm y)] &= -{\rm i} \epsilon^{ijk} \d_k \delta ({\bm x} - {\bm y}), 
\eeq
and Gauss's law ${\bm \nabla} \cdot {\bm E} = 0$; as a result, only two of the broken symmetry 
generators are independent, which correspond to the two polarizations of photons. 
In this way, massless photons with the dispersion relation $\omega=k$ and with 
two polarizations can be understood as type-A NG modes associated with the SSB 
of the 1-form symmetries \cite{Gaiotto:2014kfa, Lake:2018dqm, Hofman:2018lfz}.

\subsection{Type-B NG mode of the 1-form symmetry}
Let us turn to the original setup of QCD coupled to QED with ${\bm B}_{\rm ex}$. 
In this case, not only the 1-form electric and magnetic symmetries, but also the 
$\U(1)_{\rm A}^{\tau_3}$ symmetry is spontaneously broken in the vacuum, and so we have 
the additional corresponding symmetry broken generator,
\beq
Q_5 = \int_V n_5({\bm x})\,, \quad n_5 = \bar q \gamma_0 \gamma_5 \tau^3 q\,.
\eeq
Here $\int_V$ stands for the integral over the 3-dimensional spatial volume. 
However, $Q_5$ is not conserved in the presence of electromagnetic fields due to 
the triangle anomaly, and what is conserved instead is the combination, 
\beq
\label{tilde_Q5}
\tilde Q_5 \equiv \int_V \left(n_5 + C {\bm A} \cdot {\bm B} \right)\,.
\eeq
Here, the second term is the so-called magnetic helicity, which is gauge 
invariant under the proper boundary conditions. 

From the Lagrangian (\ref{L}), the equation of motion~(\ref{cons_E}) is modified to
\beq
\label{EOM_F}
\d_{\mu}F^{\mu \nu} = -\frac{C}{f_{\pi}} \tilde F_{\rm ex}^{\mu \nu} \d_{\mu} \pi^0\,,
\eeq
while the Bianchi identity (\ref{cons_B}) remains unchanged. Here we defined
$\tilde F_{\rm ex}^{\mu \nu} \equiv (\delta^{\mu z} \delta^{\nu 0} - \delta^{\mu 0} \delta^{\nu z}) B_{\rm ex}$.
Then, $j_{\rm M}^{\mu \nu}$ is again conserved, but $j_{\rm E}^{\mu \nu}$ itself is not. 
We can define the following conserved current by absorbing the right-hand side of 
Eq.~(\ref{EOM_F}) into the current:
\beq
\tilde j_{\rm E}^{\mu \nu} \equiv F^{\mu \nu} + \frac{C}{f_{\pi}} \tilde F_{\rm ex}^{\mu \nu} \pi^0.
\eeq

We shall now work out the commutators between the symmetry broken generators, 
$\int_{\Sigma_z} \tilde j_{\rm E}^{0z}$, $\int_{\Sigma_z} j_{\rm M}^{0z}$, and $\tilde Q_5$. 
For this purpose, we need the commutation relations for the charges 
$\tilde j_{\rm E}^{0i}$, $j_{\rm M}^{0i}$, and $n_5$:
\begin{align}
[\tilde j_{\rm E}^{0i}({\bm x}), j_{\rm M}^{0j}(\bm y)] &= {\rm i} \epsilon^{ijk} \d_k \delta ({\bm x} - {\bm y}), 
\\
\label{En5}
[\tilde j_{\rm E}^{0i}({\bm x}), n_5(\bm y)] &= -{\rm i}
C \left(2 \tilde F^{0i} - \tilde F_{\rm ex}^{0i} \right) \delta ({\bm x} - {\bm y}),
\end{align}
and the other commutators are vanishing. 
In order to derive Eq.~(\ref{En5}), we used
\beq
[\pi^0({\bm x}), n_5({\bm y})] = {\rm i} f_{\pi} \delta ({\bm x} - {\bm y}),
\eeq
and the the anomalous commutator \cite{Adler:1970qb},
\beq
[F^{0i}({\bm x}), n_5({\bm y})] = -2{\rm i}C \tilde F^{0i} \delta ({\bm x} - {\bm y}),
\eeq
which is related to the triangle anomaly in relativistic quantum field theory.
From Eq.~(\ref{tilde_Q5}) and the commutation relations (\ref{EB}), (\ref{En5}), and
\beq
\label{AE}
[A^i({\bm x}), E^j(\bm y)]= -{\rm i} \delta^{ij} \delta ({\bm x} - {\bm y}), 
\eeq
we find
\beq
\label{anom}
\frac{1}{S(\Sigma_z)} \int_{\Sigma_z} \langle[\tilde j_{\rm E}^{0z}({\bm x}), {\tilde Q_5}]\rangle = - {\rm i} C B_{\rm ex}\,, 
\eeq
where the integral is taken over the codimension-2 surface $\Sigma_z$ normal to the $z$ direction
and $S(\Sigma_z)$ is the area of $\Sigma_z$. Note that the right-hand side of Eq.~(\ref{anom}) 
originates from the second term in Eq.~(\ref{En5}) alone, since the anomalous contribution 
from the first term in Eq.~(\ref{En5}) is absorbed as the magnetic helicity into $\tilde Q_5$. 
Equation (\ref{anom}) will play a key role in what follows.

We are now ready to apply the counting rule of the NG modes in Eqs.~(\ref{NG})--(\ref{B}).
Without the external magnetic field, we have three independent symmetry broken generators, 
$\int_{\Sigma_z} \tilde j_{\rm E}^{0z}$, $\int_{\Sigma_z} j_{\rm M}^{0z}$, and $\tilde Q_5$, 
which correspond to three type-A NG modes.
In the presence of the external magnetic field, Eq.~(\ref{B}) together with the anomalous 
commutator (\ref{anom}) dictates that $N_{\rm B} = 1$, which corresponds to the NG 
mode with the quadratic dispersion in Eq.~(\ref{dispersion}). 
Since $N_{\rm BS} = 3$, Eq.~(\ref{A}) in turn leads to $N_{\rm A} = 1$, which corresponds 
to the photon with the other polarization. This is consistent with the results above, 
demonstrating that the nonrelativistic NG mode in Eq.~(\ref{dispersion}) is nothing but the 
type-B NG mode of the 1-form symmetry.%
\footnote{This should be contrasted with the nonrelativistic photon that appears in the axion 
  electrodynamics with a spatially varying and periodic $\theta$ term in the previous studies 
  \cite{Yamamoto:2015maz, Brauner:2017mui, Ozaki:2016vwu}. In those cases, the 
  quadratic dispersion relation of photons follows from the commutator (\ref{EB}) in the 
  presence of a nonvanishing $\langle {\bm \nabla} \theta \rangle$ \cite{Hidaka}, which is 
  different from our present situation. Also, the type-B NG mode appears in the 2-dimensional 
  plane perpendicular to ${\bm B}_{\rm ex}$ in our setup, while the nonrelativistic photon in 
  Refs.~\cite{Yamamoto:2015maz, Brauner:2017mui, Ozaki:2016vwu} exists only in the 
  1-dimensional line in which $\theta$ is spatially varying.} 

We remark that, because the generator for the 1-form electric symmetry, 
$\int_{\Sigma_z} \tilde j_{\rm E}^{0z}$, is restricted to the 2-dimensional surface $\Sigma_z$, 
the type-B NG mode appears only in the plane transverse to the external magnetic field. 
This is a feature specific to type-B NG modes of 1-form symmetries; in contrast, those of 
usual 0-form symmetries live in the whole 3-dimensional space.

\section{Realization in Dirac semimetals}
So far, we have focused on QCD + QED in the external magnetic field as an example. 
As is clear from the derivation, our results above are universally applicable to the chiral 
symmetry broken phase of a massless Dirac fermion coupled to a U(1) gauge field in an 
external magnetic field. 

One realization of our prediction in condensed matter systems is 3-dimensional Dirac 
semimetals, where charged massless Dirac fermions emergently appear as quasiparticles 
close to the band touching points. Since the effective coupling constant 
$\alpha_{\rm eff} \equiv \alpha/v_{\rm F}$, with $\alpha \simeq 1/137$ being the coupling constant, 
can become strong due to the smallness of the Fermi velocity $v_{\rm F} \ll 1$ in candidate 
materials for Dirac semimetals, the strong Coulomb interaction may lead to the spontaneous 
breaking of $\U(1)$ axial symmetry and the corresponding NG mode---in a way somewhat similar 
to the QCD vacuum. In fact, a gapped phase (or insulating phase) for sufficiently large 
$\alpha_{\rm eff}$ has been theoretically predicted using 
the strong-coupling expansion \cite{Sekine:2014yna, Araki:2015php}, 
Schwinger-Dyson equation \cite{Gonzalez:2015iba}, ladder approximation \cite{Gonzalez:2015tsa}, 
and lattice Monte Carlo simulations \cite{Braguta:2016vhm, Braguta:2017voo}. 
When that happens, one can show from our argument above that the type-B NG mode of the 
1-form symmetry with the quadratic dispersion relation emerges, when the external magnetic 
field is turned on. It would be interesting to search for such a new type of NG mode experimentally.

\section{Conclusion}
In this paper, we found a novel type-B NG mode of the 1-form symmetry in a theory of 
strongly interacting massless Dirac fermion coupled to U(1) gauge field in an external 
magnetic field. We have shown that the existence of this mode is dictated by the anomalous 
commutator in the symmetry algebra. 
Although we limit ourselves to systems at zero temperature in this paper, one can show, 
similarly to Ref.~\cite{Hidaka:2012ym} based on the Langevin-type low-energy effective 
theory, that this type-B NG mode appears even at finite temperature \cite{Sogabe}.

The presence of this type-B NG mode is expected to affect the low-energy dynamics 
dramatically. One example is the dynamic critical phenomenon, which is governed by 
the symmetries and low-energy degrees of freedom of a given system. 
In fact, one can show that the dynamic critical phenomenon at finite temperature 
affected by this NG mode is different from those known so far, providing a new 
dynamic universality class beyond the conventional classification. 
This will be reported elsewhere \cite{Sogabe}.

\section*{Acknowledgement}
We thank T.~Brauner, Y.~Hidaka, and R.~Yokokura for useful discussions. 
We are especially grateful to Y.~Hidaka for the explanation of the generalized global 
symmetries and for useful comments.
N.~S. is supported by JSPS KAKENHI Grant No.~17J04047. N.~Y. is supported by JSPS 
KAKENHI Grant No.~16K17703 and MEXT-Supported Program for the Strategic Research 
Foundation at Private Universities, ``Topological Science'' (Grant No.~S1511006).

\end{document}